\documentclass[preprints,article,accept,oneauthor,pdftex]{Definitions/mdpi}

\firstpage{1}
\usepackage{graphicx}
\usepackage{bm,amsfonts,amsthm,placeins}
\newcommand{\gl}[1]{(\ref{#1})}

\pubvolume{xx}
\articlenumber{5}
\pubyear{2019}
\copyrightyear{2019}
\history{Received: date; Accepted: date; Published: date}
\Title{Nonadiabatic Atomic-like State Stabilizing Antiferromagnetism
 and Mott Insulation in MnO} 
  
\Author{Ekkehard Kr\"uger \orcidA{}}

\AuthorNames{Firstname Lastname, Firstname Lastname and Firstname Lastname}

\address[1]{%
Institut f{\"u}r Materialwissenschaft, Materialphysik,
  Universit{\"a}t Stuttgart, D-70569 Stuttgart, Germany}

\corres{Correspondence: ekkehard.krueger@imw.uni-stuttgart.de}

\abstract{In this paper I report evidence that the antiferromagnetic
  and insulating ground state of MnO is caused by a nonadiabatic
  atomic-like motion as it is evidently the case in NiO. In addition,
  I show that the experimental findings of Goodwin et al.
  [Phys. Rev. Lett. (2006), 96,~047209] corroborate my suggestion that
  the rhombohedral-like distortion in antiferromagnetic MnO as well as
  in antiferromagnetic NiO is an inner distortion of the monoclinic
  base-centered Bravais lattice of the antiferromagnetic phases.}

\begin{document}
%%%%%%%%%%%%%%%%%%%%%%%%%%%%%%%%%%%%%%%%%%

%%%%%%%%%%%%%%%%%%%%%%%%%%%%%%%%%%%%%%%%%%
%\setcounter{section}{-1} %% Remove this when starting to work on the template.

\section{Introduction}
Manganese monoxide is antiferromagnetic with the N\'eel temperature
$T_N = 122K$~\cite{shull,rothI}. Above $T_N$, MnO possesses the fcc
structure $Fm3m = \Gamma^f_cO^5_h$ (225) (in parentheses I always give
the international number). Just as in antiferromagnetic NiO, the
magnetic structure of antiferromagnetic MnO is invariant under the
monoclinic base-centered magnetic group $C_c2/c$~\cite{cj} which will
be given explicitly in Eq.~\gl{eq:7}. The antiferromagnetic state is
accompanied by a small rhombohedral-like contraction of the crystal
along one of the triad axes~\cite{goodwin}, which evidently forms an
inner deformation of the monoclinic base-centered Bravais lattice
$\Gamma^b_m$~\cite{enio}. The physical origin of the rhombohedral-like
contraction in MnO is clearly the same as in NiO since the concerned
magnetic groups are the same. So, I give in Sections~\ref{sec:mgroup}
and~\ref{sec:rhombohedral} only a brief summary of Sections 3 and 4
of~\cite{enio} to make the paper selfcontained. These sections provide
the magnetic group of the antiferromagnetic state and repeat the
definition of the term ``rhombohedral-like contraction'' used in the
following sections.

In Section~\ref{sec:goodwin} the experimental findings of Goodwin et
al.~\cite{goodwin} will be interpreted. They corroborate my concept
of the rhombohedral-like contraction being an {\em inner} deformation of the
monoclinic base-centered Bravais lattice of antiferromagnetic MnO (and
NiO).

Also the other electronic features of MnO are very similar to those of
NiO: both are antiferromagnetic under the respective N\'eel
temperatures, and both are Mott insulators in the antiferromagnetic as
well as in the paramagnetic phase~\cite{austin}.  Hence, the stability
of the antiferromagnetic and the insulating state of MnO should be
caused by a nonadiabatic atomic-like motion as it is evidently the
case in NiO. In this paper I report evidence that this is true,i.e.,
that a suited nonadiabatic atomic-like motion exists also in MnO. The
starting points are the conventional band structures of paramagnetic
and antiferromagnetic MnO. In Section~\ref{sec:bandstructure} I will
explain the characteristics of the conventional band structures as
used in this paper. On the basis of the symmetry of the Bloch
functions in the band structure of MnO, I apply in
Section~\ref{sec:wannierf} the group-theoretical nonadiabatic
Heisenberg model (NHM)~\cite{enhm} to paramagnetic and
antiferromagnetic MnO. First, in Section~\ref{sec:paramagneticwf} I
will show that there exists best localized symmetry-adapted Wannier
functions at the Fermi level of paramagnetic MnO qualifying this
material to be a Mott insulator. In the following
Section~\ref{sec:antiferromagneticwf} I will show that the
nonadiabatic atomic-like motion of the electrons in antiferromagnetic
MnO stabilizes not only the antiferromagnetic state but provides also
an ideal precondition for the Mott condition to be effective in
antiferromagnetic MnO.

\section{Magnetic group of the antiferromagnetic state}
\label{sec:mgroup}
\begin{figure}[!]
\centering
\includegraphics[width=.8\textwidth]{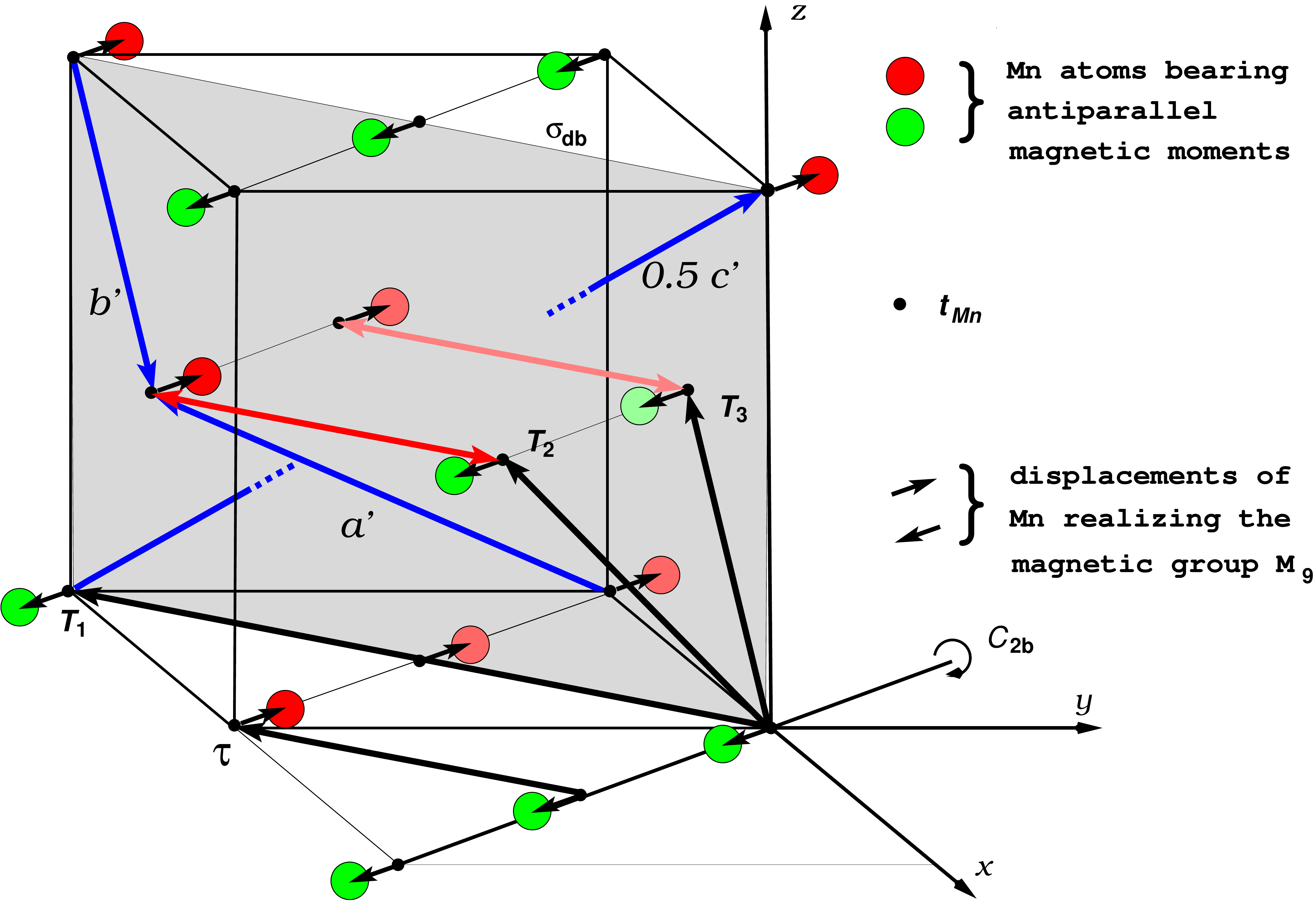}
\caption{Mangan atoms in distorted
  antiferromagnetic MnO possessing the magnetic group 
  $M_9$ (Eq.~\gl{eq:15}) based on the monoclinic base-centered
  Bravais lattice $\Gamma^b_m$. The Mn atoms
  represented by red circles bear a magnetic moment parallel or
  antiparallel to $[11\overline{2}]$ and the atoms represented by
  green circles the opposite moment. The vectors $\bm T_i$ are
  the basic translations of $\Gamma^b_m$. 
\label{fig:structures}
}
\end{figure}

This section is a brief summary of Section 3 of~\cite{enio} providing
the terms needed in Sections~\ref{sec:goodwin}
and~\ref{sec:wannierf}.  The antiferromagnetic structure of MnO is
invariant under the symmetry operations of the type IV Shubnikov
magnetic group $C_c2/c$~\cite{cj} which may be written as~\cite{bc}
\begin{equation}
  \label{eq:7}
  C_c2/c = C2/c + K\{E|\bm \tau \}C2/c,
\end{equation}
where $K$ denotes the anti-unitary operator of time-inversion,
$E$ stands for the identity operation, 
and 
\begin{equation}
  \label{eq:8}
  {\bm \tau} = \frac{1}{2}{\bm T}_1
\end{equation}
is the non-primitive translation in the group $C2/c$
as indicated in Figure~\ref{fig:structures}.

The unitary subgroup $C2/c$ (15) of $C_c2/c$ is based on the
monoclinic base-centered Bravais lattice $\Gamma^b_m$ and contains
(besides the pure translations) the 4 elements given in Equation (2)
of~\cite{enio}.  As in~\cite{enio}, I refer the magnetic group
$C_c2/c$ to as $M_{15}$ because its unitary subgroup $C2/c$ bears the
international number 15,
\begin{equation}
  \label{eq:31}
  M_{15} = C2/c + K\{E|\bm \tau \}C2/c.
\end{equation}
Though the magnetic group $M_{15}$ leaves invariant the
antiferromagnetic structure of MnO, it is not the magnetic
group of antiferromagnetic MnO because it does not possess suitable
co-representations. Instead it is the subgroup 
\begin{equation}
  \label{eq:15}
  M_9 = Cc + K\{C_{2b}|\bm 0\}Cc
\end{equation}
of $M_{15}$ allowing the system to have eigenstates with the
experimentally observed antiferromagnetic structure~\cite{enio}. The
symmetry operation $C_{2b}$ is the rotation through $\pi$ as indicated
in Figure~\ref{fig:structures}. Just as $M_{15}$, $M_9$ is based on the
monoclinic base-centered Bravais lattice $\Gamma^b_m$. The unitary
subgroup $Cc$ (9) of $M_9$ contains (besides the pure translations)
two elements,
\begin{equation}
 \label{eq:23}
Cc = \Big\{\{E|\bm 0\}, \{\sigma_{db}|\bm \tau\}\Big\}, 
\end{equation}
where $\sigma_{db}$ stands for the reflection
$\sigma_{db} = IC_{2b}$ and $I$ denotes the inversion.

Magnetoscriction alone produces the magnetic group $M_{15}$ in
MnO. Consequently, in addition to the magnetoscriction, the crystal
must be distorted in such a way that the Hamiltonian of the
nonadiabatic electron system still commutes with the elements of
$M_9$, but does not commute with the symmetry operations of
$M_{15} - M_9$.  This is achieved by exactly the one distortion of the
crystal illustrated in Figure~\ref{fig:structures}: The Mn atoms are
shifted in $\pm (\bm T_2 - \bm T_3)$ direction from their positions at
the lattice points $\bm t_{Mn}$ in Eq.~\gl{eq:1}, realizing in this
way the group $M_9$~\cite{enio}.

%%%%%%%%%%%%%%%%%%%%%%%%%%%%%%%%%%%%%%%%%%%%%%%%%%%%%%%%%%%%%%%%%%%%%
%%%%%%%%%%%%%%%%%%%%%%%%%%%%%%%%%%%%%%%%%%%%%%%%%%%%%%%%%%%%%%%%%%%%%

\section{Rhombohedral-like distortion}
\label{sec:rhombohedral}
As well as antiferromagnetic NiO, antiferromagnetic MnO is slightly
deformed by a rhombohedral-like distortion which may be interpreted as
inner distortion of the base-centered monoclinic Bravais lattice
$\Gamma^b_m$~\cite{enio}. With ``inner distortion'' I want to
emphasize that the rhombohedral-like distortion does not break the
symmetry of the magnetic group $M_9$. Figure~\ref{fig:structures}
shows the dislocations of the Mn atoms stabilizing, on the one hand,
the antiferromagnetic structure and producing, on the other hand, the
rhombohedral-like distortion, as is substantiated in~\cite{enio}.

For the sake of clarity, the basic vectors of the Bravais lattice
$\Gamma^b_m$ of $M_9$ are embedded in the paramagnetic fcc lattice of
MnO. Therefore, this Figure is also somewhat misleading because the
fcc lattice no longer exists in the antiferromagnetic phase. It is
distorted as a whole whereby the vectors ${\bm T_1}$, ${\bm T_2}$, and
${\bm T_3}$ stay basic vectors of $\Gamma^b_m$. The lattice points
${\bm t}_{Mn}$ plotted in Figure~\ref{fig:structures} are not the
positions of Mn in the fcc lattice, but are defined by the equations:
\begin{align}
\begin{split}
 \label{eq:1}
 {\bm t}_{Mn} & =  n_1{\bm T_1} + n_2{\bm T_2} + n_3{\bm T_3}\ \ \ \text{ or }\\
 {\bm t}_{Mn} & =  \frac{1}{2}\bm T_1 + n_1{\bm T_1} + n_2{\bm T_2} + n_3{\bm T_3},
\end{split}
\end{align}
where ${\bm T_1}$, ${\bm T_2}$, and ${\bm T_3}$ are the basic vectors
of $\Gamma^b_m$ and $n_1$, $n_2$, and $n_3$ are integers. Thus, the
vectors ${\bm t}_{Mn}$ are solely given in terms of the basic vectors
of $\Gamma^b_m$ detached from the paramagnetic fcc lattice.

\section{Interpretation of the experimental findings of Goodwin et
  al. }
\label{sec:goodwin}
% The base-centered monoclinic magnetic group $M_9$~\gl{eq:15} produces
% in antiferromagnetic MnO (as well as in antiferromagnetic NiO) a
% distortion closely resembling a rhombohedral distortion, see the
% detailed substantiation in~\cite{enio}.  This ``rhombohedral-like''
% structure, however, does not possess the (exact) rhombohedral magnetic
% symmetry.
Goodwin et al.~\cite{goodwin} determined the displacements
of the Mn and O atoms in the true monoclinic structure from their
positions in an assumed rhombohedral structure and gave the result in
Table I of their paper. They use the coordinates $\bf a', b'$ and
$\bf c'$ depicted in Figure~\ref{fig:structures} by the blue vectors,
depending on the translations of the base-centered monoclinic Bravais
lattice according to
\begin{eqnarray}
  \label{eq:32}
  \bm a' & = & 2\bm T_2 - \bm T_3\\\nonumber
  \bm b' & = &  -\bm T_3\\
  \bm c' & = &  -3\bm T_1 + 2\bm T_2 + 2\bm T_3,\nonumber
\end{eqnarray}
see Figure~\ref{fig:structures}. From Table I of Ref.~\cite{goodwin} we
get for the displacement $\Delta$ of a Mn1 atom
\begin{equation}
 \label{eq:33}
  \Delta = \alpha \bm a' + \beta \bm b' + \gamma \bm c'           
\end{equation}
which may be written as
\begin{equation}
  \label{eq:34}
  \Delta = -3\gamma \bm T_1 + \frac{3\alpha + \beta}{2} (\bm T_2 -
  \bm T_3) + \frac{\alpha - \beta + 4\gamma}{2} (\bm T_2 + \bm T_3),  
\end{equation}
leading with $\alpha = -2.072\cdot 10^{-3}$, $\beta = 10.11\cdot
10^{-3}$, and $\gamma = 2.708\cdot 10^{-3}$ (Table I
of~\cite{goodwin}) to
\begin{equation}
  \label{eq:35}
  \Delta = -8.124 \bm T_1 + 1.947 (\bm T_2 -
  \bm T_3) - 0.675 (\bm T_2 + \bm T_3),
\end{equation}
omitting the common factor $10^{-3}$.

The vectors $\bm T_1$ and $(\bm T_2 + \bm T_3)$ lie in the plane
related to the reflection $\sigma_{db}$, shortly referred to as ``the
plane $\sigma_{db}$'' as depicted in
Figure~\ref{fig:structures}. Hence, the first and the third summand
in~\gl{eq:35} describe a displacement of the Mn atoms parallel to the
plane $\sigma_{db}$ resulting from a modification of the translations
$\bm T_1$, $\bm T_2$, and $\bm T_3$ occurring in such a way that the
$\bm T_i$ stay basis vectors of the base-centered monoclinic Bravais
lattice~\cite{enio}. Consequently, these summands are caused by a
deformation of the crystal as a whole being still invariant under the
translations of the monoclinic lattice.

The second summand, on the other hand, describes a shift of the Mn
atoms perpendicular the the plane $\sigma_{db}$. It cannot be the
result of a modification of the lattice vectors $\bm T_i$ because such
a modification would destroy the base-centered monoclinic Bravais
lattice. These shifts of the Mn atoms in $\pm(\bm T_2 - \bm T_3)$
directions clearly identify the base-centered monoclinic magnetic
group $M_9$ as the magnetic group of antiferromagnetic MnO because
$M_9$ is the only magnetic group invariant under both the
antiferromagnetic structure and a shift of the Mn atoms perpendicular
to the plane $\sigma_{db}$~\cite{enio}.  Thus, we may interpret this
result of Goodwin et al. as follows:
\begin{enumerate}
\item The significant shifts of the Mn atoms in
  $\pm(\bm T_2 - \bm T_3)$ direction realize the magnetic group $M_9$ and
  stabilize in this way the antiferromagnetic structure, see
  Section~\ref{sec:mgroup}.
\item The observed displacements of the Mn atoms in Equation~\gl{eq:35}
  are greatest in $\bm T_1$ direction, they are even 12 times greater
  than in $(\bm T_2 + \bm T_3)$ direction. This corroborates my
  supposition~\cite{enio} that the mutual attraction between Mn atoms
  with opposite shifts in $\pm(\bm T_2 - \bm T_3)$ direction is mainly
  responsible for the rhombohedral-like deformation of the
  crystal. The displacements are maximal in $\bm T_1$ direction since
  in this direction they are parallel to the plane $\sigma_{db}$ and,
  thus, do not destroy the magnetic group $M_9$, as is illustrated by
  the red line in Figure~\ref{fig:structures}.
\end{enumerate}

Furthermore, Goodwin et al. found that the symmetry of the threefold
rotational axis implicit in a rhombohedral lattice is (slightly)
broken in MnO. This demonstrates again that we only have a
rhombohedral-like deformation rather than an (exact) rhombohedral
symmetry in antiferromagnetic MnO (just as in antiferromagnetic
NiO~\cite{enio}). On the other hand, a break of the monoclinic
base-centered symmetry with the magnetic group $M_9$ must not happen
because this would destabilize the magnetic structure of MnO.

The small dislocations of the O atoms in $\bm c'$ direction (i.e.,
parallel to $\sigma_{db}$) observed by Goodwin et al. are simply the
result of the modification of the translations $\bm T_1$, $\bm T_2$,
and $\bm T_3$ produced by the Mn atoms. The O atoms do not
actively deform the crystal.

\section{Conventional band structure}
\label{sec:bandstructure}
The conventional band structure of paramagnetic MnO in
Figure.~\ref{fig:bandstr225} is calculated by the FHI-aims (``Fritz
Haber Institute ab initio molecular simulations'') computer program
using the density functional theory (DFT)~\cite{blum1,blum2} to
compute the total energy $E_{\bm k}$ of the Bloch states in the
Brillouin zone.  By ``conventional band structure'', I mean a pure
one-electron band-structure not taking into account any correlation
effects. The strong correlation effects responsible for the stable
antiferromagnetic state and the nonmetallic behavior of MnO enter by
the postulates~\cite{enhm} of the group-theoretical NHM defining a
strongly correlated nonadiabatic atomic-like motion at the Fermi
level.

The NHM uses only a more qualitative run of the energy $E_{\bm k}$ in
the band structure. It is the symmetry of the Bloch states in the
points of symmetry of the Brillouin zone which characterizes an energy
band within the NHM. The FHI-aims program uses spherical harmonics as
basis functions and provides the possibility of an output of the
special linear combinations of these functions used at a point
$\bm k$. Thus, I was able to write a C++ program to determine the
symmetry of the Bloch functions at the points of symmetry in the
Brillouin zone using the symmetry of the spherical harmonics as given
in \cite{bc}.

\section{Symmetry-adapted and optimally localized Wannier functions in
  MnO}
\label{sec:wannierf}
The NHM defines in narrow, partly filled electronic energy bands a
strongly correlated nonadiabatic atomic-like motion stabilized by the
nonadiabatic condensation energy $\Delta E$ defined in Equation (2.20)
of~\cite{enhm}. This atomic-like motion is not represented by atomic
or hybrid atomic functions, but strictly by symmetry-adapted and
optimally localized Wannier functions forming an (exact) unitary
transformation of the Bloch functions of the energy band under
consideration~\cite{enhm,theoriewf}. When the band is roughly
half-filled and one of narrowest bands in the band structure, the
special symmetry and spin-dependence of these Wannier functions
qualifies a material to be superconducting~\cite{josn,ebi,theoriewf},
magnetic~\cite{ea,ef,bamn2as2,theoriewf}, or a Mott
insulator~\cite{bamn2as2,enio}. In the following
Subsection~\ref{sec:paramagneticwf} I will show that paramagnetic MnO
possesses optimally localized Wannier functions including all the
electrons at the Fermi level and adapted to the fcc symmetry of the
paramagnetic phase. These functions define a nonadiabatic atomic-like
motion evidently responsible for the Mott insulation. In the second
Subsection~\ref{sec:antiferromagneticwf} I will show that the
electrons near the Fermi level of antiferromagnetic MnO can be
represented by Wannier functions comprising all the electrons at the
Fermi level and adapted to the magnetic group $M_9$ of the
antiferromagnetic phase. The nonadiabatic atomic-like motion
represented by these Wannier functions evidently stabilizes both the
antiferromagnetic state and the Mott insulation.

\subsection{Wannier functions symmetry-adapted to the paramagnetic fcc
  structure}
\label{sec:paramagneticwf}

%%%%%%%%%%%%%%%%%%%%%%%%%%%%%%%%%%%%%%%%%%%%%%%%%%%%%%%%%%%%%%%%%%%%%
 \begin{figure*}[h]
 \includegraphics[width=.95\textwidth,angle=0]{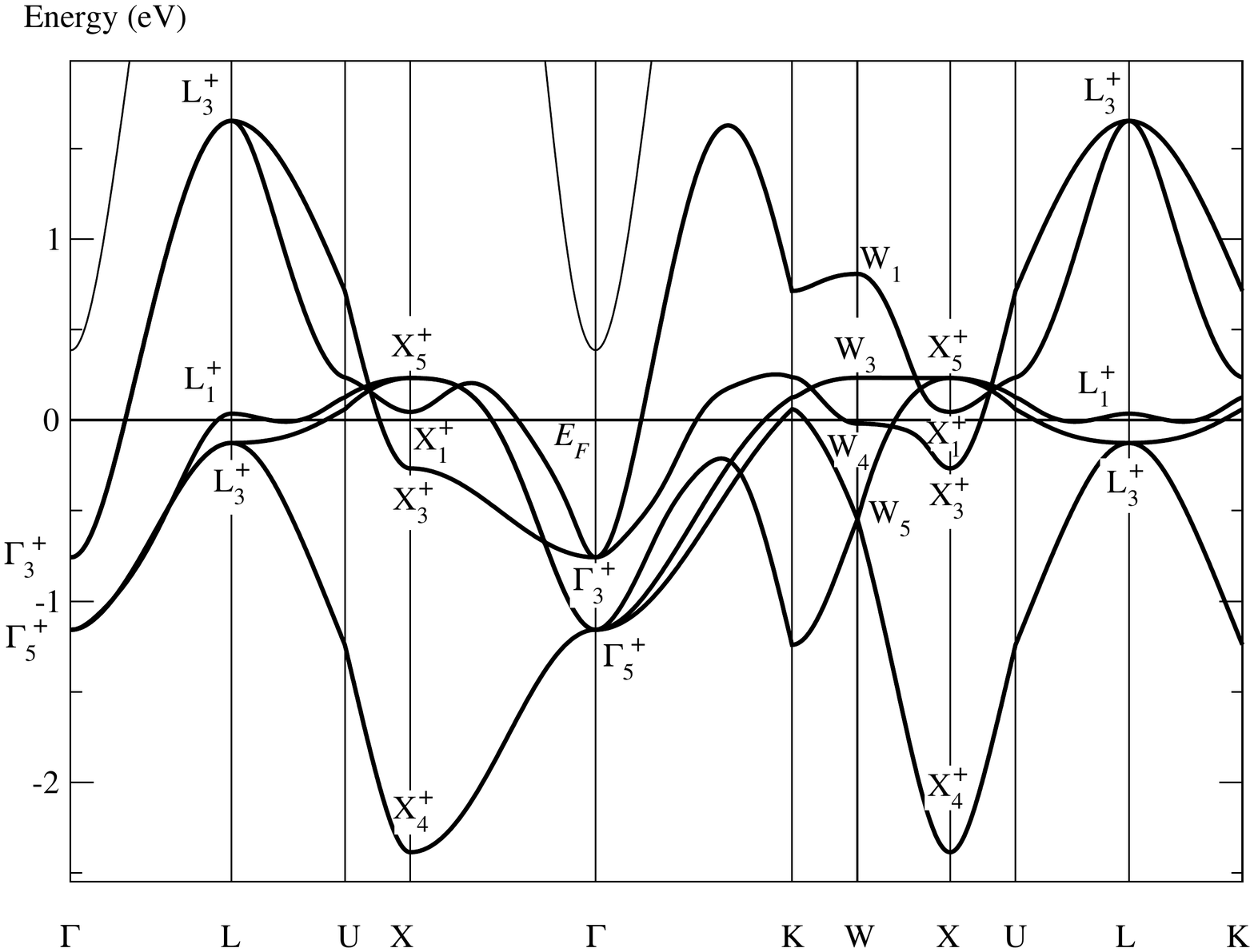}%
 \caption{Conventional (Section~\ref{sec:bandstructure}) band structure of
   paramagnetic fcc MnO as calculated by the FHI-aims program
   \cite{blum1,blum2}, using the length $a = 4.426A$ of the
   fcc paramagnetic unit cell given in
Ref.~\cite{shull}, with symmetry labels determined by the
author. The notations of the points of symmetry in the Brillouin zone
for $\Gamma^f_c$ follow Figure 3.14 of Ref.~\cite{bc} and the symmetry
labels are defined in
Table A1 of~\cite{enio}. The  band highlighted by the bold
lines forms an insulating band of $d$ symmetry consisting of 5 branches.
 \label{fig:bandstr225}
}
 \end{figure*}
%%%%%%%%%%%%%%%%%%%%%%%%%%%%%%%%%%%%%%%%%%%%%%%%%%%%%%%%%%%%%%%%%%%%%

 All the information we need in this section is given in
 Figure~\ref{fig:bandstr225} and
 Table~\ref{tab:wf_225}. Figure~\ref{fig:bandstr225} shows the
 conventional band structure of paramagnetic MnO and
 Table~\ref{tab:wf_225} is an excerpt of Table 1 of~\cite{eniopara}.
 While Table 1 of~\cite{eniopara} lists $all$ the optimally localized
 symmetry-adapted Wannier functions in MnO centered at the Mn or O
 atoms and adapted to the paramagnetic space group $Fm3m$ (225),
 Table~\ref{tab:wf_225} lists only the two bands with Wannier
 functions of $\Gamma^+_3$ and $\Gamma^+_5$ symmetry centered at the
 Mn atoms. The bands have two and three branches, respectively,
 yielding together five Wannier functions at each Mn atom.

 %%%%%%%%%%%%%%%%%%%%%%%%%%%%%%%%%%%%%%%%%%%%%%%%%%%%%%%%%%%%%%%%%%%%%%

\begin{table}[t]
\caption{
Symmetry labels of the Bloch functions at the points of symmetry in
the Brillouin zone for $Fm3m$ (225) of two energy bands with
symmetry-adapted and optimally localized Wannier functions centered at
the Mn atoms 
\label{tab:wf_225}}
\begin{center}
\begin{tabular}[t]{cccccc}
 & Mn($000$) & $\Gamma$ & $X$ & $L$ & $W$\\
\hline
Band 5 & $\Gamma^+_3$ & $\Gamma^+_3$ & $X^+_1$ + $X^+_3$ & $L^+_3$ & $W_1$ + $W_4$\\
Band 8 & $\Gamma^+_5$ & $\Gamma^+_5$ & $X^+_4$ + $X^+_5$ & $L^+_1$ + $L^+_3$ & $W_3$ + $W_5$\\
\hline\\
\end{tabular}%\hspace{1cm}
\end{center}
% \ \\
\begin{flushleft}
Notes to Table~\ref{tab:wf_225}
\end{flushleft}
\begin{enumerate}
\item The notations of the points of symmetry in the Brillouin zone
  for $\Gamma^f_c$ follow Figure 3.14 of Ref.~\cite{bc}, and the
  symmetry notations of the Bloch functions are defined in Table A1
  of~\cite{enio}.
\item The bands are determined following Theorem 5
of~\cite{theoriewf}, cf. Section 2 of~\cite{eniopara}.
\item The point group $G_{0Mn}$ of the positions~\cite{theoriewf} of
  the Mn atoms are equal to the full cubic point group $O_h$.  The
  Wannier functions belong to the representation of $G_{0Mn}$ included
  below the atom.
\end{enumerate}
\end{table}
 
 %%%%%%%%%%%%%%%%%%%%%%%%%%%%%%%%%%%%%%%%%%%%%%%%%%%%%%%%%%%%%%%%%%%%%%

By inspection of Figure~\ref{fig:bandstr225} and
Table~\ref{tab:wf_225} we recognize that the closed band highlighted
by the bold lines in the band structure of paramagnetic MnO is
characterized by the symmetry
\begin{equation}
  \label{eq:2}
  \Gamma^+_3 + \Gamma^+_5,\ L^+_3 + L^+_1 + L^+_3,\ X^+_5 + X^+_1 +
  X^+_3 + X^+_4,\ W_1 + W_3 + W_4 + W_5  
\end{equation}
of the two bands in Table~\ref{tab:wf_225} with Wannier functions
centered at the Mn atoms. The set of these five Wannier functions has
$d$ symmetry since an atomic $d$ orbital splits into a
$\Gamma_3^+$ and a $\Gamma_5^+$ state in the fcc crystal field, see
Table 2.7 of~\cite{bc}. On the other hand, no partly filled band in
the band structure of paramagnetic MnO possesses the symmetry of any
band in Table~1 b of~\cite{eniopara}, listing the bands with Wannier
functions centered at the O atoms. Hence, the energy band~\gl{eq:2}
originates entirely from a $d$ orbital of Mn. This $d$ band is an
insulating band qualifying paramagnetic MnO to be a Mott insulator
because it consists of all the branches crossing the Fermi
level~\cite{eniopara}.

\subsection{Wannier functions symmetry-adapted to the antiferromagnetic structure}
\label{sec:antiferromagneticwf}

%%%%%%%%%%%%%%%%%%%%%%%%%%%%%%%%%%%%%%%%%%%%%%%%%%%%%%%%%%%%%%%%%%%%%%%% 
 \begin{figure*}[!]
%\begin{minipage}[t]{.65\textwidth}
 \includegraphics[width=.95\textwidth,angle=0]{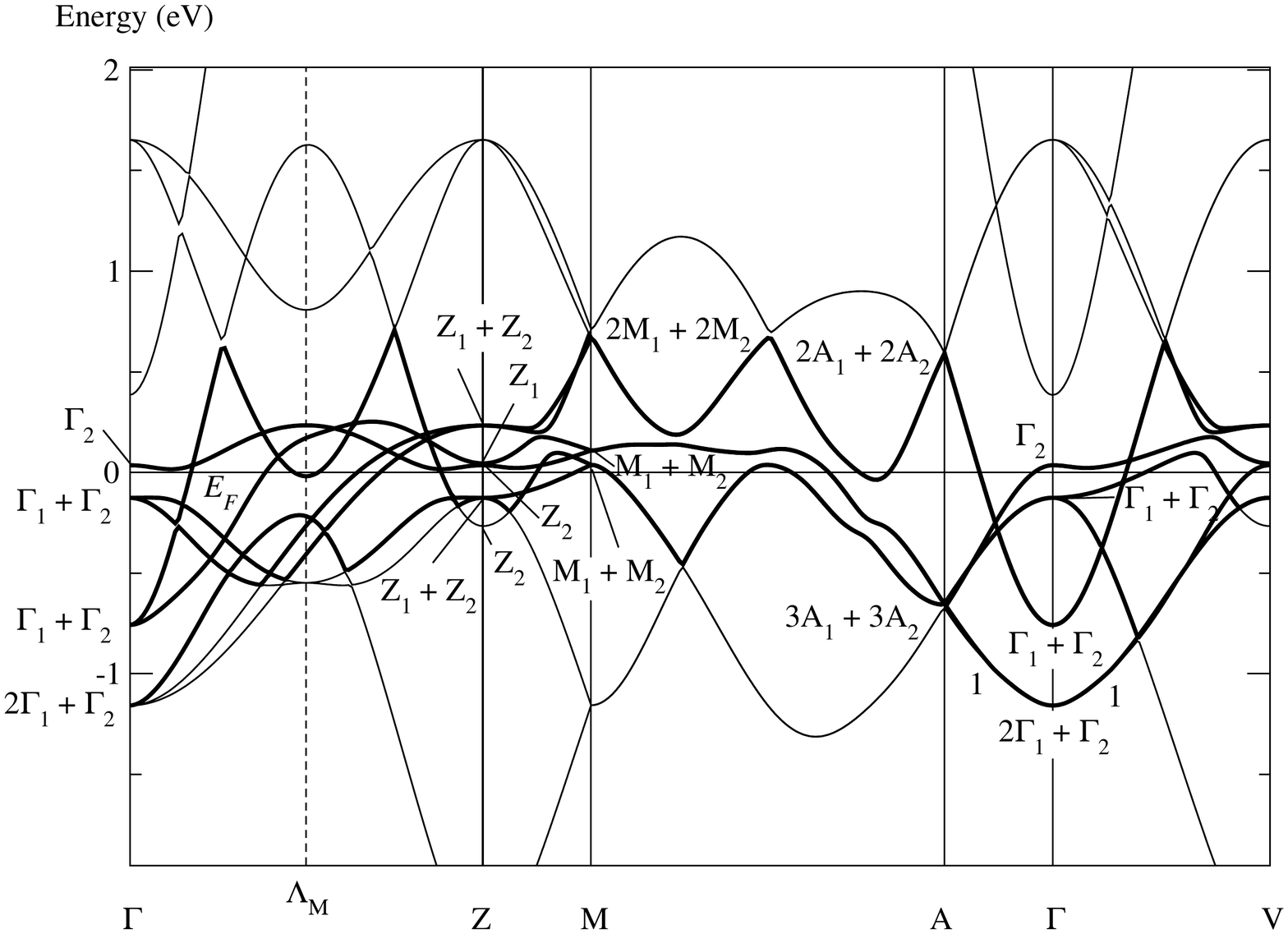}%
%\end{minipage}
 %\vspace{.52cm}
 \caption{ The band structure of MnO given in
   Figure~\ref{fig:bandstr225} folded into the Brillouin zone for the
   monoclinic base centered Bravais lattice $\Gamma^b_m$ of the
   magnetic group $M_9$. The band highlighted by the bold lines forms
   a magnetic super band consisting of six branches assigned to
   the manganese or oxygen atoms.  The symmetry labels are defined in
   Table A4 of~\cite{enio} and are determined from
   Figure~\ref{fig:bandstr225} by means of Table A5 of~\cite{enio}.  The
   notations of the points of symmetry follow Figure~3.4~of
   Ref.~\cite{bc}. The midpoint $\Lambda_{\text{M}}$ of the line
   $\overline{\Gamma Z}$ is equivalent to the points
   $W' (\overline{\frac{1}{4}}\frac{1}{4}\frac{1}{2})$ and
   $\Sigma' (\frac{1}{4}\overline{\frac{1}{4}}0)$ in the Brillouin
   zone for the paramagnetic fcc lattice. The number ``1'' on the
   line $\overline{A\Gamma V}$ indicates that here only one branch
   belongs to the magnetic band.
\label{fig:bandstr9}
}
 \end{figure*}
%%%%%%%%%%%%%%%%%%%%%%%%%%%%%%%%%%%%%%%%%%%%%%%%%%%%%%%%%%%%%%%%%%%%%%%%

%%%%%%%%%%%%%%%%%%%%%%%%%%%%%%%%%%%%%%%%%%%%%%%%%%%%%%%%%%%%%%%%%%%%%%%%
\begin{table}[!]
\caption{
Symmetry labels of the Bloch functions at the points of symmetry in
the Brillouin zone for $Cc$ (9) of all
the energy bands with symmetry-adapted and optimally  
localized Wannier functions centered at the Mn $\big($Table (a)$\big)$ and O
$\big($Table (b)$\big)$ atoms, respectively. 
\label{tab:wf_9}}
\begin{flushleft}
\begin{tabular}[t]{cccccccccc}
(a)\ \ {\bf Mn} & Mn$_1(000)$ & Mn$_2(\overline{\frac{1}{2}}00)$ & $K\{C_{2b}|\bm
0\}$ & $\Gamma$ & $A$ & $Z$ & $M$ & $L$  & $V$\\
\hline
Band 1 & $\bm{d}_{1}$ & $\bm{d}_{1}$ & OK & $\Gamma_1$ + $\Gamma_2$ & $A_1$ + $A_2$ & $Z_1$ + $Z_2$ & $M_1$ + $M_2$ & 2$L_1$ & 2$V_1$\\
\hline\\
\end{tabular}\hspace{1cm}
\begin{tabular}[t]{cccccccccc}
(b)\ \ {\bf O} & O$_1(\overline{\frac{1}{4}}\frac{1}{2}\overline{\frac{1}{2}})$ & O$_2(\overline{\frac{3}{4}}\frac{1}{2}\overline{\frac{1}{2}})$ & $K\{C_{2b}|\bm
0\}$ & $\Gamma$ & $A$ & $Z$ & $M$ & $L$ & $V$\\
\hline
Band 1 & $\bm{d}_{1}$ & $\bm{d}_{1}$ &  OK  & $\Gamma_1$ + $\Gamma_2$ & $A_1$ + $A_2$ & $Z_1$ + $Z_2$ & $M_1$ + $M_2$ & 2$L_1$ & 2$V_1$\\
\hline\\
\end{tabular}\hspace{1cm}
\end{flushleft}
%\ \\
%\ \\
\begin{flushleft}
Notes to Table~\ref{tab:wf_9}
\end{flushleft}
\begin{enumerate}
\item The notations of the points of symmetry in the Brillouin zone
  for $\Gamma^b_m$ follow Figure 3.4 of Ref.~\cite{bc}
  and the symmetry notations of the Bloch functions are defined in
  Table A4 of~\cite{enio}.
\item The bands are determined following Theorem 5
  of~\cite{theoriewf}.
\item The point groups $G_{0Mn}$ and $G_{0O}$ of the
positions~\cite{theoriewf} of the Mn respective O atoms 
contain, in each case, only the identity operation:
  \begin{equation}
    \label{eq:24}
    G_{0Mn} = G_{0O} = \Big\{ \{E|\bm 0\} \Big\}.
    \end{equation}
    Thus, the Wannier functions at the Mn or O atoms belong to the
    simple representation
\begin{center}
\begin{tabular}[t]{cc}
%\multicolumn{2}{c}{Mn, O}\\
 & $\{E|\bm 0\}$\\
\hline
$\bm{d}_{1}$ & 1\\
\hline\\
\end{tabular}\hspace{1cm}
\end{center}
of $G_{0Mn}$ and $G_{0O}$.
\item The symmetry of band 1 in Table~\ref{tab:wf_9} (a) coincides
  fully with the symmetry of band 1 in Table~\ref{tab:wf_9} (b).
\item Each row defines a band consisting of two branches with Bloch
  functions that can be unitarily transformed into Wannier functions
  being:
\begin{itemize}
\item as well localized as possible; 
\item centered at the Mn$_1$ and Mn$_2$ (Table (a)) or O$_1$ and O$_2$
  (Table (b)) atoms; and
\item symmetry adapted to the unitary subgroup $Cc$ (9) of the
  magnetic group $M_9$ according to
  \begin{equation}
    \label{eq:25}
    \begin{array}{lcl}
    P(\{\sigma_{db}|\bm \tau\})w_{Mn_1}(\bm r) & = & w_{Mn_2}(\bm r), \\
    P(\{\sigma_{db}|\bm \tau\})w_{Mn_2}(\bm r) & = & w_{Mn_1}(\bm r), \\
    P(\{\sigma_{db}|\bm \tau\})w_{O_1}(\bm r) & = & w_{O_2}(\bm r),  \\ 
    P(\{\sigma_{db}|\bm \tau\})w_{O_2}(\bm r) & = & w_{O_1}(\bm r);
  \end{array}
\end{equation}
  where $w_{Mn_1}(\bm r), w_{Mn_2}(\bm r), w_{O_1}(\bm r)$, and
  $w_{O_2}(\bm r)$ denote the Wannier functions centered at the Mn and
  O atoms, respectively. 
  \item The entry ``OK'' indicates that the Wannier functions follow
  also Theorem 7 of Ref.\ \cite{theoriewf} and, consequently, may be chosen
  symmetry-adapted to the magnetic group $M_9$. Thus, in
  addition to Equation~\gl{eq:25}, we have:  
  \begin{equation}
    \label{eq:26}
    \begin{array}{lcl}
    KP(\{C_{2b}|\bm 0\})w_{Mn_{1}}(\bm r) & = & w_{Mn_{1}}(\bm r),\\
    KP(\{C_{2b}|\bm 0\})w_{Mn_{2}}(\bm r) & = & w_{Mn_{2}}(\bm r),\\
    KP(\{C_{2b}|\bm 0\})w_{O_{1}}(\bm r) & = & w_{O_{2}}(\bm r), \\
    KP(\{C_{2b}|\bm 0\})w_{O_{2}}(\bm r) & = & w_{O_{1}}(\bm r),
    \end{array}
  \end{equation}
\end{itemize}
cf. the more detailed notes to Table A2 in~\cite{enio}.  
\end{enumerate}
\end{table}
%%%%%%%%%%%%%%%%%%%%%%%%%%%%%%%%%%%%%%%%%%%%%%%%%%%%%%%%%%%%%%%%%%%%%%%
All the information we now need is given in Figure~\ref{fig:bandstr9}
and in Table~\ref{tab:wf_9}. This table is a shortened copy of Table
A2 of~\cite{enio} omitting the detailed notes to this Table
in~\cite{enio} (of course, these notes are also valid for
Table~\ref{tab:wf_9} in this paper when ``Ni'' is replaced by
``Mn''). Table~\ref{tab:wf_9} lists all the optimally localized
Wannier functions centered at the Mn or O atoms and symmetry-adapted
to the magnetic group $M_9$. In both cases, there exists only one band
with conciding symmetries.

When folding the band structure of paramagnetic MnO given in
Figure~\ref{fig:bandstr225} into the Brillouin zone of the
monoclinic-base centered magnetic structure, we receive the band
structure plotted in Figure~\ref{fig:bandstr9}. The narrow band
highlighted by the bold lines comprises six branches with the symmetry
\begin{equation}
  \label{eq:3}
  3\Gamma_1 + 3\Gamma_2,\ 3Z_1 + 3Z_2,\ 3M_1 + 3M_2,\ 3A_1 + 3A_2.
\end{equation}
Thus, band 1 (with two branches) listed in Table~\ref{tab:wf_9} (a) as
well as band 1 in Table~\ref{tab:wf_9} (b) exists three times in the
band structure of antiferromagnetic MnO. Hence, we may unitarily
transform the Bloch functions of this band into optimally localized
Wannier functions symmetry adapted to the magnetic group $M_9$. In
doing so, we first have four possibilities: we may generate either six
Wannier functions centered at the two Manganese atoms (with three
Wannier functions at Mn$_1$ and three Wannier functions at Mn$_2$),
four Wannier functions centered at the manganese atoms and two Wannier
functions centered at the oxygen atoms, two Wannier functions centered
at the manganese atoms and four Wannier functions centered at the
oxygen atoms, or six Wannier functions centered at the oxygen atoms in
the unit cell. The first possibility does certainly not lead to a
stable atomic-like state as consequence of the Coulomb repulsion
between localized states at the same position, and the last
possibility has no physical significance because it does not provide
localized states at the Mn atoms which bear the magnetic moments. In
the remaining two cases the highlighted band is a magnetic band. It
provides localized Wannier states allowing the electrons to perform a
nonadiabatic atomic-like motion stabilizing a magnetic state with the
magnetic group $M_9$~\cite{enio,ea}. Moreover, it is a magnetic super
band since it comprises all the branches crossing the Fermi level and,
consequently, qualifying MnO to be a Mott insulator also in the
antiferromagnetic phase~\cite{enio}.

\section{Results}
The paper is concerned with three striking features of MnO:
\begin{enumerate}
\item The insulating ground state of both paramagnetic and
  antiferromagnetic MnO,
\item the stability of the antiferromagnetic state,
\item the rhombohedral-like deformation in the antiferromagnetic phase,
\end{enumerate}
characterizing likewise the isomorphic transition-metal monoxide NiO.
The aim of this paper was to show that these striking electronic
properties of MnO have the same physical origin as in NiO: Just as in
NiO~\cite{enio},
\begin{itemize}
\item the antiferromagnetic state in MnO is evidently stabilized by
  the nonadiabatic atomic-like motion in a magnetic band. This
  magnetic band is even a magnetic super band because it comprises all
  the electrons at the Fermi level. Thus, the special atomic-like
  motion in this band qualifies antiferromagnetic MnO to be a Mott
  insulator.
\item the Bloch functions of a (roughly) half-filled energy band in
  the paramagnetic band structure of MnO can be unitarily transformed
  into optimally localized Wannier functions symmetry-adapted to the
  fcc symmetry of the paramagnetic phase. These Wannier functions are
  situated at the Mn atoms, have $d$ symmetry and comprise all the
  electrons at the Fermi level. Thus, the atomic-like motion
  represented by these Wannier functions qualifies also paramagnetic
  MnO to be a Mott insulator.
\item the magnetic structure is stabilized by a shift
  of the Mn atoms in $\pm(\bm T_2 - \bm T_3)$ direction. These
  shifts evidently produce the rhombohedral-like deformation of
  the crystal because the attraction between the Mn atoms increases
  slightly when the Mn atoms are shifted in opposite directions.  This
  concept presented in~\cite{enio} is corroborated by the experimental
  observations of Goodwin et al.~\cite{goodwin}.
\item the rhombohedral-like distortion does not possess a rhombohedral
  (trigonal) space group but is an inner distortion of the
  base-centered monoclinic magnetic group $M_9$ in
  Equation~\gl{eq:15}.  The group $M_9$, on the other
  hand, must not be broken because it stabilizes the antiferromagnetic
  structure.
\end{itemize}

%%%%%%%%%%%%%%%%%%%%%%%%%%%%%%%%%%%%%%%%%%
\section{Discussion}
\label{sec:discussion}
The results of this paper demonstrate once more that the nonadiabatic
atomic-like motion defined within the NHM has physical reality. So
they confirm my former findings suggesting that
superconductivity~\cite{josn,ebi,theoriewf},
magnetism~\cite{ea,ef,bamn2as2,theoriewf} and Mott
insulation~\cite{enio,eniopara,bamn2as2} are produced by the nonadiabatic
atomic-like motion defined within the NHM.

In addition, the paper corroborates my former
suggestion~\cite{ea,lafeaso1,enio} that we can determine by group
theory whether or not a magnetic state with the magnetic group $M$
may be an eigenstate in a system invariant under time inversion.
\vspace{6pt} 

%%%%%%%%%%%%%%%%%%%%%%%%%%%%%%%%%%%%%%%%%%
%% optional
%\supplementary{The following are available online at \linksupplementary{s1}, Figure S1: title, Table S1: title, Video S1: title.}


%%%%%%%%%%%%%%%%%%%%%%%%%%%%%%%%%%%%%%%%%%
\funding{This publication was supported by the Open Access Publishing Fund of the University of Stuttgart.}
%%%%%%%%%%%%%%%%%%%%%%%%%%%%%%%%%%%%%%%%%%
\acknowledgments{I am very indebted to Guido Schmitz for his support
  of my work.}

%%%%%%%%%%%%%%%%%%%%%%%%%%%%%%%%%%%%%%%%%%
\conflictsofinterest{The author declares no conflict of interest.}
%%%%%%%%%%%%%%%%%%%%%%%%%%%%%%%%%%%%%%%%%%
%% optional
\abbreviations{The following abbreviations are used in this manuscript:\\

\noindent 
\begin{tabular}{@{}ll}
NHM & Nonadiabatic Heisenberg model\\
$E$ & Identity operation\\
$I$ & Inversion\\
$C_{2b}$ & Rotation through $\pi$ as indicated in Figure~\ref{fig:structures}\\
$\sigma_{db}$ & Reflection $IC_{2b}$\\
$K$ & anti-unitary operator of time inversion
\end{tabular}}

%%%%%%%%%%%%%%%%%%%%%%%%%%%%%%%%%%%%%%%%%%
%% optional
\appendixtitles{no} %Leave argument "no" if all appendix headings stay EMPTY (then no dot is printed after "Appendix A"). If the appendix sections contain a heading then change the argument to "yes".

\bibliographystyle{mdpi}
%%%%%%%%%%%%%%%%%%%%%%%%%%%%%%%%%%%%%%%%%%
\reftitle{References}

\end{document}